\documentclass[journal]{IEEEtran}
\ifCLASSINFOpdf
\else
\fi
\usepackage{multirow}
\usepackage{graphicx}
\usepackage{subfigure}
\usepackage{amsmath}

\begin{document}
%
\title{Coarse2Fine: Two-layer Fusion for Image Retrieval}
%
%
%

\author{le~Dong,~
        Gaipeng~Kong,~
        Wenpu~Dong,~
        Liang~Zheng~
        and~Qi~Tian
\thanks{M. Shell was with the Department
of Electrical and Computer Engineering, Georgia Institute of Technology, Atlanta,
GA, 30332 USA e-mail: (see http://www.michaelshell.org/contact.html).}
\thanks{J. Doe and J. Doe are with Anonymous University.}
\thanks{Manuscript received April 19, 2005; revised August 26, 2015.}}

%
%

\markboth{Journal of \LaTeX\ Class Files,~Vol.~14, No.~8, August~2015}%
{Shell \MakeLowercase{\textit{et al.}}: Bare Demo of IEEEtran.cls for IEEE Journals}
%



\maketitle

\begin{abstract}
This paper addresses the problem of large-scale image retrieval. We propose a two-layer fusion method which takes advantage of global and local cues and ranks database images from coarse to fine (C2F). Departing from the previous methods fusing multiple image descriptors simultaneously, C2F is featured by a layered procedure composed by filtering and refining. In particular, C2F consists of three components. 1) Distractor filtering. With holistic representations, noise images are filtered out from the database, so the number of candidate images to be used for comparison with the query can be greatly reduced. 2) Adaptive weighting. For a certain query, the similarity of candidate images can be estimated by holistic similarity scores in complementary to the local ones. 3) Candidate refining. Accurate retrieval is conducted via local features, combining the pre-computed adaptive weights. Experiments are presented on two benchmarks, \emph{i.e.,} Holidays and Ukbench datasets. We show that our method outperforms recent fusion methods in terms of storage consumption and computation complexity, and that the accuracy is competitive to the state-of-the-arts.

\end{abstract}


%
\IEEEpeerreviewmaketitle

\section{Introduction}
%
%
%
%
\IEEEPARstart{T}{his} paper considers the task of accurate image retrieval on a large scale. Given a query image, we aim at finding the  similar images from the database. A number of retrieval models have been proposed in this scenario, among which local feature based model, \emph{i.e.}, bag-of-words (BOW) \cite{sivic2003video}, has obtained widespread applications. Nevertheless, traditional BOW is lack of spatial layout and may encounter quantization errors. Besides, local feature based models usually lead to huge storage consumption, which is limited in real-world applications featured by the rapid expansion of the image volume. On the contrary, holistic representations such as HSV represent an image via global vectors and obtain good scalability, while the retrieval accuracy is less desirable.

\begin{figure}[t]
\begin{center}
\includegraphics[width=1\linewidth]{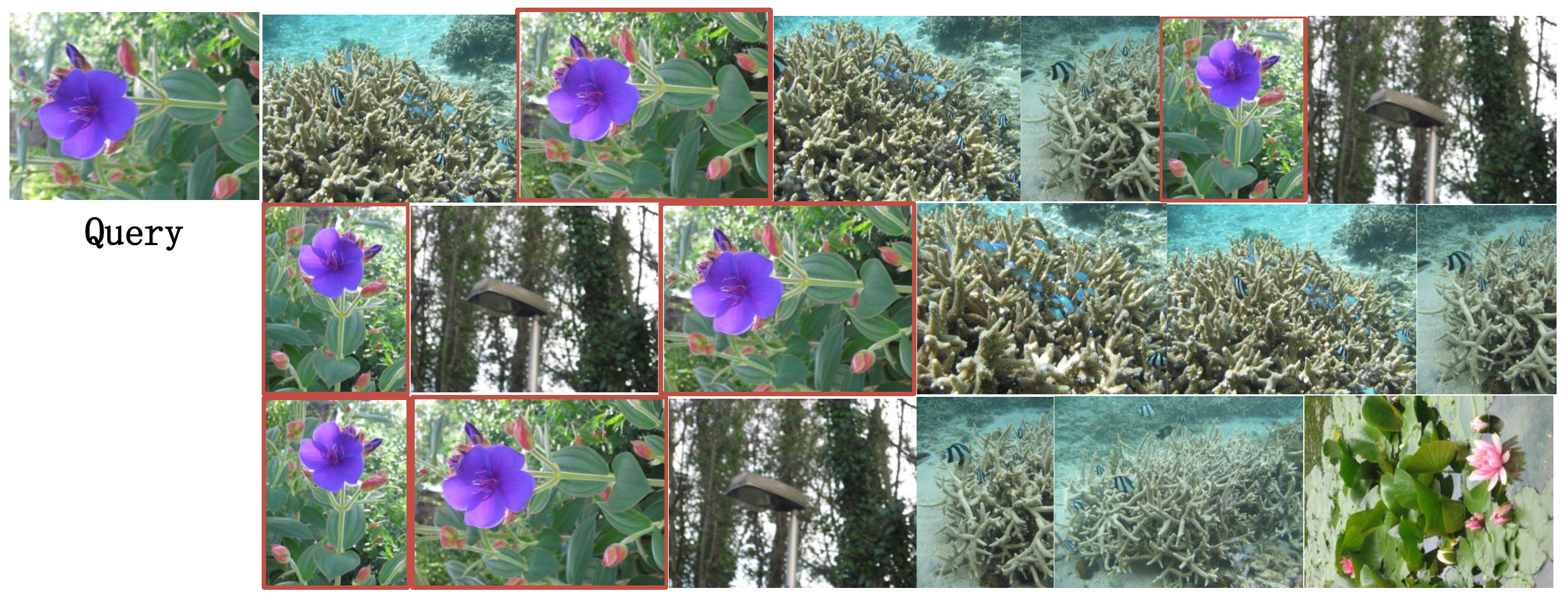}
\end{center}
\caption{A sample query from the Holidays dataset and its retrieval results obtained by holistic representation (first row), BOW (second row) and the proposed two-layer fusion framework (third row). }
\label{Fig: example}
\end{figure}

Generally speaking, local descriptors and holistic representations demonstrate distinct strengths in finding similar images. Local features are capable of capturing local image patterns or textures, while holistic representations delineate overall feature distributions in images. Thus, a variety of methods \cite{zhang2012query,zhang2013semantic,zheng2014packing,zheng2015query,liu2014visual} are proposed to integrate the strengths of local and holistic features to yield more satisfying retrieval results.

Zhang \emph{et al.} \cite{zhang2012query} propose a graph based fusion approach to merge and rerank multiple retrieval sets. Although this method achieves desirable accuracy, there are some disadvantages,\emph{ e.g.}, being not robust to the dynamic changes of the dataset or being unable to find reciprocal neighborhoods. Zheng \emph{et al.} \cite{zheng2014packing} propose a coupled Multi-Index framework to perform feature fusion at indexing level. Specifically, color and SIFT features are coupled into a multi-dimensional inverted index to filter out false positive SIFT matches. The \emph{c-MI} seems to achieve higher accuracy accompanied with less query time, while the storage consumption is still remarkable when the image dataset rapidly expands. Another representative work is the query-adaptive late fusion method proposed in \cite{zheng2015query}. To a certain query, \cite{zheng2015query} extracts five different image descriptors and each of them is used to search the whole dataset independently. The initial score curves obtained via these representations are normalized by the references and later used to evaluate the effectiveness of the corresponding descriptors. This query-adaptive method does have competitive retrieval performance, while it heavily relies on the number of to-be-fused features and needs considerable amount of storage to save these features, as well as the reference curves.

Different from the above fusion methods utilizing multiple image descriptors to search the entire database simultaneously, C2F proposed in this paper is a two-layer fusion procedure working from coarse to fine. The first stage aims at filtering out distractors from the dataset, which helps enhance the retrieval efficiency and reduce the memory consumption to be encountered in the next step. The subsequent operation focuses on refining the rank list from the filtered candidates, which further improves the retrieval accuracy (see Figure \ref{Fig: example} for an illustration).

To be more specific, we firstly exploit the holistic descriptors to filter out distractor images that are dissimilar to the query from the database, and thus the number of candidate images used to compared with the query can be greatly reduced. This process is efficient and the retrieved candidates usually appear globally similar. Then, according to the similarity scores obtained via holistic features, we acquire $K$ candidate images which are more similar to the query when compared with other images in the database. Moreover, for each candidate image, we design an adaptive weight by holistic similarity scores to measure its similarity degree to the query. Finally, accurate image search is conducted on candidate image pool via local features. In this step, adaptive weights of the candidate images are used in refining the initial retrieval list.

The main contribution of the proposed framework consists in the coarse-to-fine fusion of retrieval sets given by different methods, which has two merits: 1) by filtering out distractors from the database, the number of candidate images to be used for comparison with the query can be greatly reduced. 2) given a certain query, the relative weight of each candidate can be automatically evaluated via global similarity scores, which helps ensure the stability of the retrieval performance. We validate the performance of C2F on two public datasets, \emph{i.e.}, Ukbench and Holidays datasets. The evaluation shows that our method compares favorably with the recent state of the art.

The remainder of the paper is organized as follows: Section 2 highlights the related works; We describe the formulation details of C2F in Section 3; Section 4 provides comprehensive experimental results to validate the superiority of C2F; Finally, Section 5 concludes the paper and states directions for future work.

\section{Related Work}

A variety of works have been exposed to improve the image retrieval performance. In this section, we briefly introduce several closely related methods.

To obtain a discriminative image representation with local features, BOW based methods are usually adopted. In BOW model, a codebook is generated off-line by unsupervised clustering algorithms, such as k-means, hierarchical k-means \cite{nister2006scalable}, and approximate k-means \cite{philbin2007object}. Each local feature is quantized to one or a few visual words by Approximate Nearest Neighbor algorithms. Then the image is represented by a high-dimensional histogram over the visual codebook. This process is usually accompanied by the quantization error \cite{philbin2008lost,zheng2014bayes}.

To correct quantization defects, hard assignment can be replaced with schemes such as soft assignment \cite{philbin2008lost}, multiple assignment \cite{jegou2008hamming}, sparse coding \cite{yang2009linear}, \emph{etc}. Nevertheless, quantizing a single feature to several visual words \cite{philbin2008lost, jegou2008hamming, yang2009linear} will introduce more storage burden and higher search complexity. Representatively, \cite{jegou2008hamming} proposes to generate a ${D_b}$-bit binary signature to each local feature with the hashing approach via a smaller codebook, for example 20,000. The true matches are defined as those local features that are not only quantized to the same visual word but also have small Hamming distance between their binary signatures.

\begin{figure*}[t]
\begin{center}
\includegraphics[width=1\linewidth]{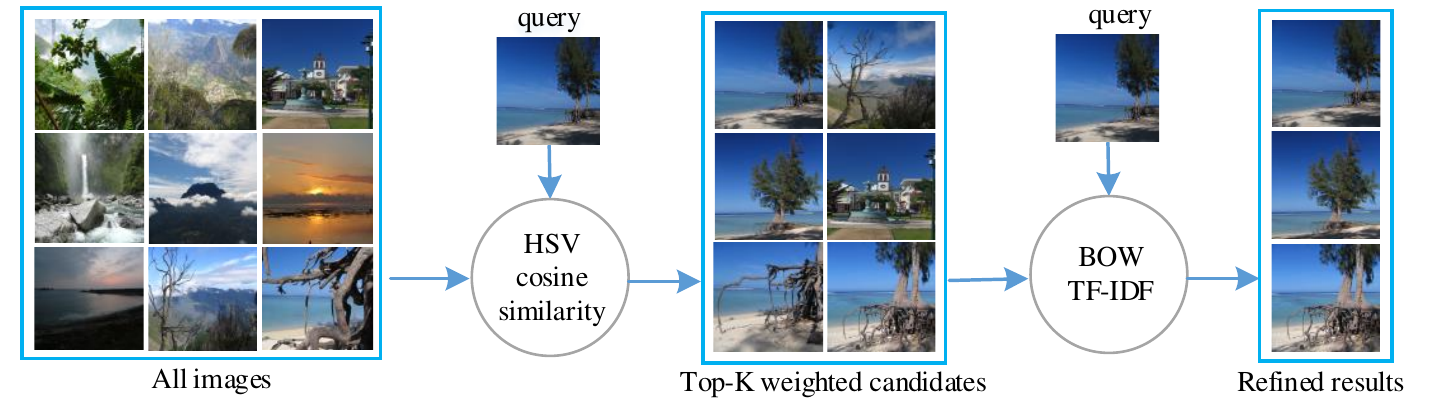}
\end{center}
\caption{C2F via coupling filtering and refining. 1) Given a query, holistic representation (HSV histograms) is used to filter out distractors from all images. 2) Top-$K$ candidates with adaptive weights are selected via global similarity scores. 3) Accurate retrieval is conducted on candidates via local features(BOW), combining the effectiveness of weights.}
\label{Fig: framework}
\end{figure*}

On the other hand, \cite{Liang2014Compact} proposes a compact feature based clustering (CFC) to represent images. \cite{Zhou2015Deep} improves the retrieval performance by automatically learning robust visual features and hash functions. \cite{Dong2015Unsupervised} proposes a discriminative light unsupervised learning network to learn a compact image representation. \cite{dong2016holons} employs a two-layer hierarchical scheme to extract the global information and statistics of the local feature set of image datasets. There is no doubt that methods based on compact holistic features are efficient in computation and memory usage, as well as more suitable to large-scale dataset. Since holistic features tend to be less invariant than local features and more sensitive to image transformations, their retrieval precision is often lower compared to local feature based methods.

As a consequence, it would be desirable if one method can effectively combine the complementary cues of local and holistic features. Rank aggregation \cite{fagin2003efficient} is a solution to fuse local and holistic cues at the rank level, but the effectiveness of rank aggregation is discounted when there is no intersection among the top candidates retrieved by the local and holistic feature based methods, which does occasionally occur. To overcome this weakness, \cite{zhang2012query} proposes an undirected-graph based fusion approach to further enhance the retrieval precision. Built on this undirected-graph, \cite{liu2014visual} proposes a directed graph which is robust to outlier distraction. \cite{Das2015Query} propose a graph-based optimization framework to leverage category independent object proposals  for logo search in a large scale image database. \cite{zhang2013semantic} proposes a semantic-aware co-indexing algorithm to jointly embed local invariant features and semantic attributes into the inverted indexes. \cite{wengert2011bag} introduces a simple color signature generation procedure and embeds local color features into the inverted index to provide color information by Bag of Colors. In \cite{zheng2014packing}, a couple Multi-Index framework is proposed to perform feature fusion at indexing level by couple complementary features into a multi-dimensional inverted index. \cite{zheng2015query} proposes a late fusion at score level. Particularly, the effectiveness of each to-be-fused feature is estimated in an unsupervised, query-adaptive manner. \cite{Yalong2014Bag} proposes a bag-of-words based deep neural network for image retrieval task, which learns high-level image representation and maps images into bag-of-words space.

From the above analysis, it can be seen that considerable efforts have been made to improve the retrieval accuracy via fusion schemes, such as graph-based \cite{zhang2012query, liu2014visual}, query-adaptive late fusion \cite{zheng2015query}. Usually, with more heterogeneous features, higher accuracy can be obtained, while more information needs to be stored and higher computational complexity incurred, correspondingly. Therefore, how to maintain a good balance between accuracy and complexity is an important issue when focusing on large-scale data. Different from the above fusion methods, in this work, we propose a two-layer fusion framework to rank database images from coarse to fine (C2F). This method improves retrieval efficiency and accuracy reliably.

\section{Two-layer fusion framework}

In this section, we present the detailed structure of the proposed C2F aiming at accurate image retrieval on large-scale databases. C2F is a two-layer fusion framework which consists of three key components. 1) It filters distractors from the whole database with holistic representations, and thus the number of images to be used for comparison with the query can be greatly reduced. 2) For a certain query, we design an adaptive weight for each selected candidate image according to the similarity scores obtained via holistic features. 3) We further refine the candidate set with local features, considering the impact of adaptive weights. The mechanism of C2F is presented in Figure \ref{Fig: framework}. In the next subsections, we will elaborate each component of the framework.

\subsection{Distractor filtering via holistic representations} \label{Method: choose}

We assume that there are a total of $N$ color images in the database $\left\{ X_n\right\}_{n=1}^N$. Using holistic features, such as HSV, each database image is represented as a $P$-dimensional histogram vector $V_n=\left\{h_1, h_2, \ldots , h_{P} \right\}$, and the image set can be defined as $\left\{V_1, V_2, \ldots , V_N \right\}$. In order to reduce the impact of bins with large values, HSV histograms of both query and dataset images are l1-normalized and square scaled, similar to the rootSIFT \cite{arandjelovic2012three}:
 \begin{align}
 & h_i = \frac{h_{i}}{\sqrt{\sum_{i=1}^{P}h_{i}}}, \\
 & h_i = |h_{i}|^{\alpha},
 \end{align}
where $\alpha$ is a coefficient and its value is set to 0.5.

To select candidate images from the dataset, we adopt the cosine distance between vectors of the query and the database images as similarity measurement:
\begin{align}
s_{q,d}^G = cos(q,d) = \frac{\sum_{i=1}^{P}V_{q,i}\cdot V_{d,i}}{\sqrt{\sum_{i=1}^{P}V_{q,i}^2}\sqrt{\sum_{i=1}^{P}V_{d,i}^2}},
\end{align}
where $q$ is the query, $d$ is a database image, and $s_{q,d}^G$ represents the similarity score obtained by global feature. The cosine similarity between the query and all the dataset images can be defined as:
\begin{align}
s_{q}^G = (s_{q,1}, s_{q,2}, \ldots , s_{q,N})^G.
\end{align}
Particularly, a larger cosine score corresponds to higher similarity with the query.

Note that there exists large variationa among the database image. Thus, we hold the opinion that most database images have significant differences from the query. Hence, the holistic image representation can be used to quickly filter out these noise images. For each query, $K$ candidates that share similar information with the query are adaptively selected from the original database. Different queries commonly have different candidate images.

\subsection{Adaptive weights for candidates} \label{Method: weight}

After the filtering stage, $K$ candidate images more similar to the query are selected from the original image database. The similarity of the query with different candidates varies extensively. It is important to design an adaptive weight for each candidate image to evaluate its similarity level with the query. Considering that holistic representations, \emph{i.e.}, HSV histograms, delineate overall color feature distributions in images, the cosine values computed in Sec. \ref{Method: choose} are used as the basis of learning weights. In particular, the similarity scores of the query with $K$ candidates are represented as $(s_{q,1}, s_{q,2}, \ldots , s_{q,K})^G$.

The motivation behind the weighting method is that the larger cosine values correspond to higher importance of the candidate images. To overcome the dictatorship incurred by the candidates with larger scores, the cosine similarity values of the corresponding candidate images are min-max normalized to $\left [ 0,1 \right ]$:

\begin{align}
& s_{q,i}^G = (\frac{s_{q,i} - min (s_{q,i})}{max (s_{q,i}) - min (s_{q,i})})^G.
\end{align}

Then, the normalized scores are used as the adaptive weights of the corresponding $K$ candidate images.

\begin{align}
& s_{q,i}^G = (\frac{s_{q,i}}{\sum_{i=1}^{K}s_{q,i}})^G, \\
& w(q,i) = s_{q,i}^G, i = 1, 2, \ldots , K.
\end{align}

Experimental results show that adaptive weights significantly improves the accuracy of image retrieval on both Holidays and Ukbench datasets. The performance of C2F with and without wights are compared in Table \ref{Fig: weights}.

\subsection{Further refinement with local features} \label{method: accurate}

Given the $K$ candidate images and their corresponding weights of a certain query, we focus on utilizing local features to find out ground truths.

Firstly, we extract SIFT descriptors from an image by Hessian-affine detector \begin{math} I=\left [ x_1, x_2,\ldots,x_M \right] \in R^{D\times M}\end{math}. The image contains $M$ local descriptors in $D$ dimensions.

Then, a codebook trained on Filckr60k dataset by Approximate K-Means (AKM) \cite{philbin2007object} clustering algorithm is used to quantize SIFT features. Specifically, each local feature $x_i$ of an image is assigned to its nearest visual word of the codebook, represented by the ID of the corresponding visual word.

\begin{equation}
q(x_i)=arg \min ||x_i-\mu_j||^2, ~~\mu_j\in C.
\end{equation}

Next, to enhance the discriminative validity of the visual words, a $d_b$-bit binary signatures feature is generated for each SIFT feature \begin{math} b(x) = \left [ b_1(x), b_2(x), \ldots, b_{d_b}(x) \right ] \end{math}, which encodes the location of the SIFT descriptor within the Voronoi cell. The distance between two descriptors $x$ and $y$ lying in the same cell is approximated by the Hamming distance of their corresponding binary signatures.
\begin{align}
\label{gongshi10}
h(b(x), b(y)) = \sum_{i=1}^{d_b}\left | b_i(x) - b_i(y) \right |.
\end{align}

At this point, a SIFT descriptor is represented by $q(x)$ and $b(x)$, and a standard inverted index is generated. Then the matching function is defined as:
\begin{equation}
\begin{split}
f(x,y) =
\begin{cases}
 1 ,~~ if ~~ q(x)=q(y)~~ and
\\ ~~~~~~h(b(x),b(y))\leq h_t,\\
0,~~~~ otherwise
\end{cases}
\end{split}
\end{equation}
where $h$ is the Hamming distance defined in E.q. \ref{gongshi10}, and $h_t$ is a fixed hamming threshold.

The similarity score between the query and the candidate image is:
\begin{align}
\begin{split}
s_{q,d}^L = \sum_{j=1}^{M} tf~idf(q(x))^2 ~~~ if q(x)=q(y)~~and \\
~~~~~~~~~~~~~~~~~~ h(b(x),b(y))\leq h_t,
\end{split}
\end{align}
where $M$ is the number of local feature in the query image $q$, $x$ and $y$ is the local feature of the query and dataset image, respectively, L represents that the score is computed via local features.

Considering the similarity degrees measured by holistic representations, the sorted list of the $K$ candidate images obtained via local features are refined according to the adaptive weights learned in Sec. \ref{Method: weight}. The final similarity scores of the query and the candidate images are defined as:
\begin{align}
s_{q,d}^F = s_{q,d}^L\times w_{q,d},
\end{align}
where $s_{q,d}^F$ is the final similarity score between the query and the dataset image d, $w_{q,d}$ is the weight obtained by the intuitively or steadily weighted method.

Overall, we propose a two-layer fusion framework which takes advantage of global and local cues and ranks database images from coarse to fine. Considering the efficiency and low storage consumption of holistic features, we use the HSV histograms to coarsely search the original database and filter out noise images. Then, accurate image search is conducted on candidate images, combining the adaptive weights learned via global similarity scores to further enhance the retrieval accuracy. Experimental results in Sec. \ref{exper: state-of-the-arts} demonstrate that the retrieval accuracy of C2F is competitive to state of the art.

\subsection{Complexity and Scalability} \label{method: analysis}

In order to illustrate the following analysis more clearly, we provide a brief description of these parameters: $|N|$ is the number of dataset images, $|Q|$ is the number of query images, $|K|$ is the number of candidate images, and $|F|$ is the number of features to-be-fused.

The rising processing time of the image retrieval mainly caused by the increasing number of candidate images, which are used for comparison with the query. In this paper, We propose a two-layer fusion method which takes advantage of global and local cues and ranks database images from coarse to fine. The total comparison number in C2F is $O(|N+K|)$, which is a relative small value compared with other fusion methods \cite{zhang2012query,zheng2015query}. For example, the comparison number in \cite{zheng2015query} is $O(|FN|)$, especially, $|F|$ is set to 5 in the experiment. Correspondingly, the shorten of the comparisons in C2F will certainly contribution to the efficiency and the storage consumption. The memory storage consumed by C2F is quite smaller than that of \cite{zheng2015query}. This is because \cite{zheng2015query} need to save multiple features of all images of the database, while C2F stores holistic features of all database images and local features of only $K$ candidate images.

\begin{table*}[!htbp]
\centering
\caption{The filtering and refining effect of C2F. $K$ is the number of candidates selected via holistic representation. We use the HSV histograms and BOW to filter out distractors and refine candidates, respectively. In particular, no weights is embed in C2F.}
\label{table: filter}
\begin{tabular}{|c|cccccccccccc|} \hline
Holidays ($K$) & 0 & 100 & 200 & 300 & 400 & 500 & 600 & 700 & 800 & 900 & 1000& 1491\\ 
HSV+HE & 61.95 & 79.81 & 80.65 & 80.68 & 80.73 & 81.23 & 81.55 & 82.11 & 82.27& 82.25 & 82.25 & 80.16 \\ \hline \hline
UKbench ($K$) & 0 & 200 & 400 & 600 &800 &1000 &1200& 1400 &1600& 1800&2000 & 10200 \\ 
HSV+HE & 3.19 & 3.621 & 3.623 & 3.623 & 3.621 & 3.617 & 3.614 &3.609 & 3.606 & 3.604 & 3.603	& 3.484\\ \hline
\end{tabular}
\end{table*}

The computational cost incurred by the proposed two-layer fusion framework is quite small. C2F computes the adaptive weighs via holistic similarity scores, this processing time is less than 1 second for over a million database images in the experiments. The weighted method used in C2F is straightforward when compared with the normalization method in \cite{zheng2015query}. And compared with the graph fusion method \cite{zhang2012query} constructing a weighted graph offline for each query from the retrieval results of one method, C2F does not need any offline processing scheme and can automatically adapt to the dynamic changes in the database.

\section{Experiments}

In following sections, we describe the datasets and the evaluation protocols (Sec. \ref{exper: dataset}), the baseline features (Sec. \ref{exper: baseline}) used in C2F, the analysis on experimental results (Sec. \ref{exper: results}), the efficiency and storage of C2F (Sec. \ref{exper: efficiency}), and the discussion on C2F (Sec. \ref{exper: discuss}).

\subsection{Datasets and Evaluation Protocol} \label{exper: dataset}

We evaluate our framework on two commonly used benchmarks, \emph{i.e.}, the Holidays and Ukbench datasets. With Holidays, we add 1M distractor images from ImageNet \cite{deng2009imagenet} for large-scale experiments.

\emph{\textbf{Holidays \cite{jegou2008hamming}}} dataset is composed of 1,491 personal holidays images undergoing various transformations. There are 500 queries in total and the other 991 images are the similar ones corresponding to queries. Generally, most queries have 2 - 3 ground truth images. As for the evaluation mechanism, we adopt Mean Average Precision (mAP) \cite{philbin2007object} to calculate retrieval accuracy. It is the mean value of Average Precision (AP), which computes the area under the precision-recall curve for each query. Specifically, precision is defined as the ratio of retrieved positive images to the total number retrieved and recall is the ratio of the number of retrieved positive images to the total number of positive images in the dataset. The ideal precision-recall curve has precision of 1.

\emph{\textbf{Ukbench \cite{nister2006scalable}}} dataset contains 10,200 images which is made up of 2,550 objects for each has four images taken from different viewpoints. In the experiments, all the 10,200 images are used as query in turn. The expected results of using each query to retrieve the Ukbench database are the four images with the same object. In term of evaluation manner, N-S score is used to calculate the retrieval accuracy.
The accuracy of each query is measured by counting the number of correct images in the top-4 ranked images of returned retrieval results. Then the retrieval performance on Ukbench is averaged over all test queries (N-S score is 4 maximum).

\textbf{\emph{ImageNet 1M}}
ImageNet \cite{deng2009imagenet} is an image data\-base organized according to the WordNet hierarchy, in which each node of the hierarchy is depicted by hundreds or thousands of images. Since ImageNet dataset contains sufficient images with large variations and is readily accessible, it is well suitable to benchmark the accuracy, computation, and memory usage for the large-scale image retrieval. In the experiments, we use 1 million images of 1,000 categories in ImageNet as distractors to evaluate the scalability of C2F. Following the experimental setting principles of \cite{philbin2007object, yang2011object}, we use the original 500 queries and the corresponding ground truths of Holidays as the queries and ground truths of the large-scale dataset. We assume there is no ground truth in the newly added images form ImageNet. mAP is used to measure the retrieval accuracy.

\subsection{Baseline Features} \label{exper: baseline}

Holistic representation and local features are used in the C2F to enhance the retrieval performance, especially, we consider the baseline of HSV histograms \cite{zheng2015query} and BOW \cite{jegou2008hamming}, as well as some recent technologies to improve retrieval accuracy.

\begin{table} [!htbp]
\centering
\caption{Retrieval accuracy with baseline features.}
\label{table: baseline}
\begin{tabular}{|c|c|c|} \hline
Datasets & HSV & BOW \\ \hline
Ukbench (N-S score)  & 3.19 & 3.484 \\ \hline
Holidays (mAP(\%))  & 61.95 & 80.16 \\ \hline
\end{tabular}
\end{table}

\emph{\textbf{HSV}} We compute a 1,000-dimensional histogram for each image. The parameters of bins for H, S, V are 20, 10, 5, respectively.

\emph{\textbf{BOW}} Following the implementation setup in \cite{jegou2008hamming}, scale-invariant keypoints are detected with the Hessian-affine \cite{philbin2007object} detector. For each SIFT feature, we use the quantization scheme as in \cite{philbin2007object}. To improve recall, rootSIFT \cite{arandjelovic2012three} and Multiple Assignment (MA) are applied. Moreover, complying with \cite{zheng2015query}, we use 128-bit Hamming signature with the Hamming threshold and weighting parameter set to 52 and 26, respectively. Retrieval accuracy of the baseline features is presented in Table \ref{table: baseline}.

\subsection{Experimental results} \label{exper: results}

In this following , we make a detailed analysis on the performance of C2F with different implementation details.

\subsubsection{Retrieval performance of C2F} \label{exper: C2F}

\emph{\textbf{Filtering.}} In order to evaluate the performance of the filtering mechanism, we conduct experiments by filtering out different numbers of distractors. In particular, noise images are discarded via HSV, and accurate image retrieval is further conducted on the selected candidates via local features. Besides, to better illustrate the effectiveness of the filtering and refining, retrieval accuracies elaborated in Table 2 are implemented without adaptive weights.

Table \ref{table: filter} presents the performance of C2F on Holidays and Ukbench dataset with different elimination degrees. For example, on Holidays, $K$ = 200 represents that 1291 images are regarded as distractors and discarded, while $K$ = 1491 means that retrieval experiments are performed on all dataset images via BOW. When the number of candidates is set to 800, the retrieval result of the C2F is 82.27, which is 1.09\% higher than that obtained by retrieving the whole dataset with local features (mAP = 80.16). The filter performance of the C2F is also well verified on Ukbench. The N-S score of the C2F with 1000 candidates is 0.133 higher than that of 10200 images (no image is filtered out from Ukbench). Experimental results demonstrate that filtering out distractor images is quite important to improve the final retrieval performance.

\emph{\textbf{Refining.}} We also conduct complementary experiments to evaluate the performance of the refining strategy, namely, whether the candidates selected from image dataset are further refined via local features. On both Holidays and Ukbench, $K$ = 0 means that no candidates are used to further enhance the retrieval results computed via HSV histogram.

Experimental results in Table \ref{table: filter} indicate that further refining on candidates can significantly enhance the retrieval accuracy at the cost of a little more memory consumption. For example, on Holidays dataset, the accuracy of the C2F with 100 candidates is 18.6\% higher than that of K = 0. And this growth will be more apparent when more candidates are selected, \emph{e.g.}, the accuracy of the C2F with 500 candidates is 85.11. The effectiveness of the refining on Ukbench is consistent with that on Holidays. When 200 candidates are selected to refine the retrieval results obtained via HSV (N-S = 3.19), search accuracy is increased to 3.693. When K is set to 1000, we get a desirable accuracy (N-S = 3.725). In general, the experimental results on both Holidays and Ukbench demonstrate that the filtering and refining strategies adopted in C2F can significantly enhance the retrieval accuracy in a simple but effective manner.

\begin{figure}[!htbp]
\begin{center}
\includegraphics[width=1.0\linewidth]{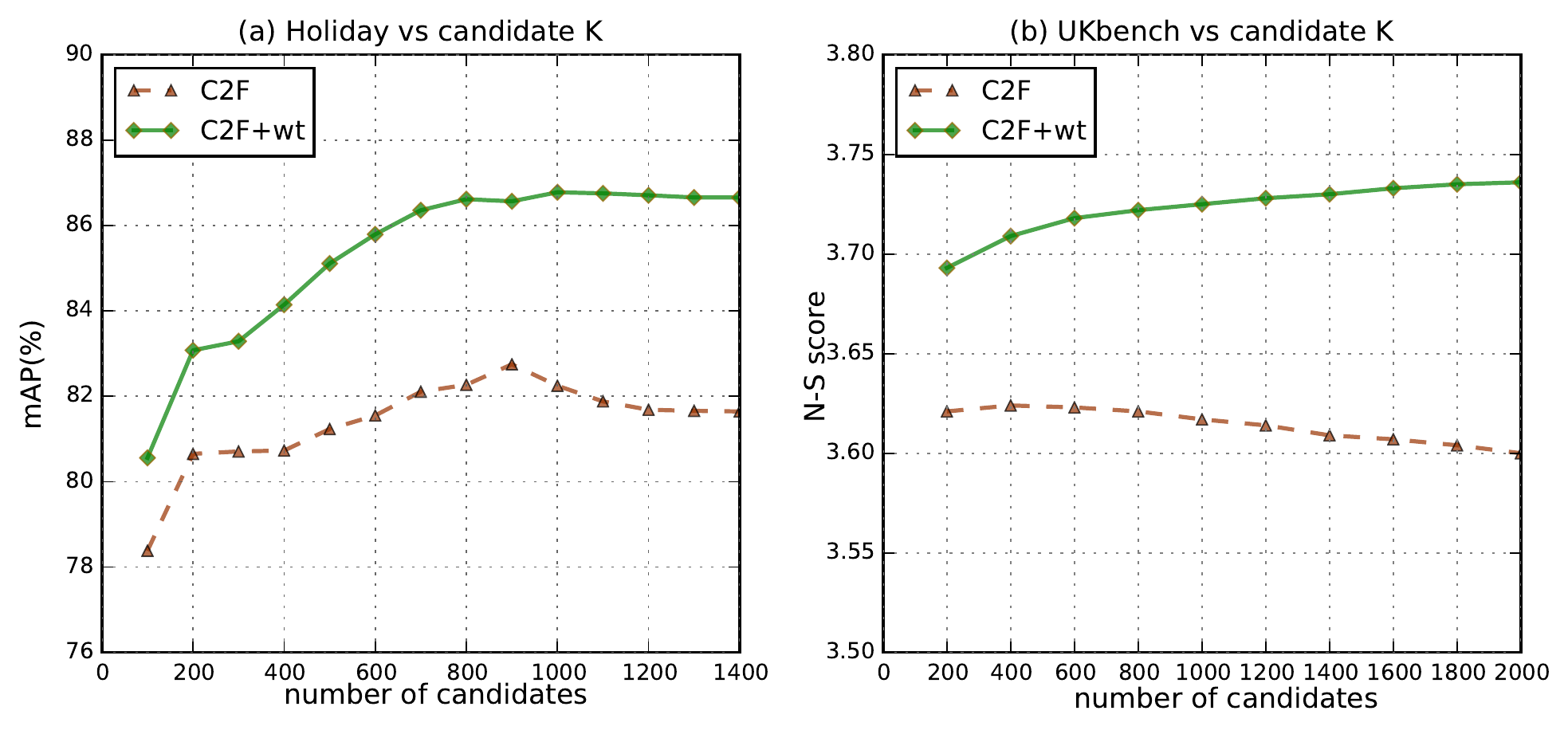}
\end{center}
\caption{Sensitivity to candidate $K$ on (a) Holidays and (b) Ukbench datasets. We use the HSV histograms to filter out distractors and BOW to further refine candidates. }
\label{Fig: candidates}
\end{figure}

\subsubsection{Impact of the candidate number} \label{exper: candidates}

In this section, we conduct experiments to validate the performance of C2F with different number of candidates. In particular, since Holidays only contains 1491 images in total, we set the number of candidates $K$ vary from 100 to 1000. While for the Ukbench dataset, we vary $K$ from 200 to 2000.

From Figure \ref{Fig: candidates}, we observe that on both Holidays and Ukbench datset, the retrieval accuracy improves when the candidate number increases. More candidates will certainly help improve the retrieval performance, however, such increase is not always guaranteed. When all ground truths of one query are recalled, the retrieval performance shows subtle improvement. This is because more candidates mean more distractors. We attribute such phenomenon to the negative impact of noisy images, that is, some noisy images will be selected when the $K$ is set to a large value. The noisy images will contribute nothing and even drag down the final retrieval performance.

\begin{table*} [!htbp]
\centering
\caption{Performance comparison with the state-of-the-art.}
\label{table: arts}
\begin{tabular}{|c|c c c c c c c c c c c c c c|} \hline
Methods & Ours & \cite{zheng2015query} & \cite{zheng2014packing}&\cite{zheng2014bayes} &\cite{zhang2013semantic} &\cite{deng2013visual} &\cite{tolias2013aggregate} &\cite{qin2013query}&\cite{shen2012object}&\cite{zhang2012query} &\cite{wang2011contextual}&\cite{wengert2011bag}&\cite{qin2011hello}&\cite{jegou2010improving} \\ \hline \hline
\texttt{Ukbench}  & 3.737 & \textbf{3.755} & 3.71 & 3.62 & 3.60 & 3.75 &-  &-  & 3.52 & 3.77 & 3.56 & 3.50 & 3.67 &3.42 \\
\texttt{Holidays}   & \textbf{86.78} & 84.47 & 84.0 & 81.9 & 80.86 & 84.7 &82.2 &82.1 & -  & 84.6 & 78.1 & 78.9 & 42.3 &81.3 \\ \hline
\end{tabular}
\end{table*}

\begin{table*} [!htbp]
\centering
\caption{Memory consumption (MB) and the average query time on holidays against the number of candidates. Feature extraction and quantization time is excluded.}
\label{table: memory}
\begin{tabular}{|l|cccccccccc|} \hline
Candidates & 100 & 200 & 300 & 400 & 500 & 600 & 700 & 800 & 900 & 1000 \\ \hline
Memory (MB) & 17.24 & 33.82 & 50.21 & 66.45 & 82.67 & 98.91 & 115.26 & 131.92 & 149.00 & 166.30 \\ \hline
Time (s) & 1.14 & 1.99 & 2.78 & 3.49 & 4.83 & 5.66 & 6.88 & 7.99 & 9.14 & 10.43\\ \hline
\end{tabular}
\end{table*}

\begin{table*} [!htbp]
\centering
\caption{Memory consumption and average query time on Ukbench against the number of candidates.}
\label{table: efficiency}
\begin{tabular}{|l|cccccccccc|} \hline
Candidates & 200 & 400 & 600 & 800 & 1000 & 1200 & 1400 & 1600 & 1800 & 2000\\ \hline
Memory (MB) & 17.69 & 34.80 & 52.13 & 69.65 & 87.34 & 105.14 & 123.16 & 141.26 & 159.49 & 177.93\\ \hline
Time (s) & 0.27 & 0.51 & 0.65 & 1.03 & 1.28 & 1.56 & 1.98 & 2.13 & 2.34 & 2.56 \\ \hline
\end{tabular}
\end{table*}

From the Figure \ref{Fig: candidates}(a), it can be observed that, on Holidays dataset, the number of candidates $K$ = 1000 obtains the best retrieval accuracy among all other values. With the adaptive weights, the mAP of candidates with 800 is 1.67\% higher than that with 500 (mAP = 85.11). When the number of candidate continues to grow, the retrieval accuracy starts to drop slightly. For example, the mAPs of candidate-1200 and candidate-1400 are 0.07\% and 0.09\% lower than that achieved by candidate-1000, respectively. Another noteworthy phenomenon is that the mAP with the least candidates ($K$ = 100) is lowest, \emph{e.g.} mAP = 80.56 with adaptive weights, this accuracy loss is caused by the improper filtering. In other words, the ground truths corresponding to some certain queries are filtered out due to the small number of candidate images.

\begin{figure} [!htbp]
\begin{center}
\includegraphics[width=1.0\linewidth] {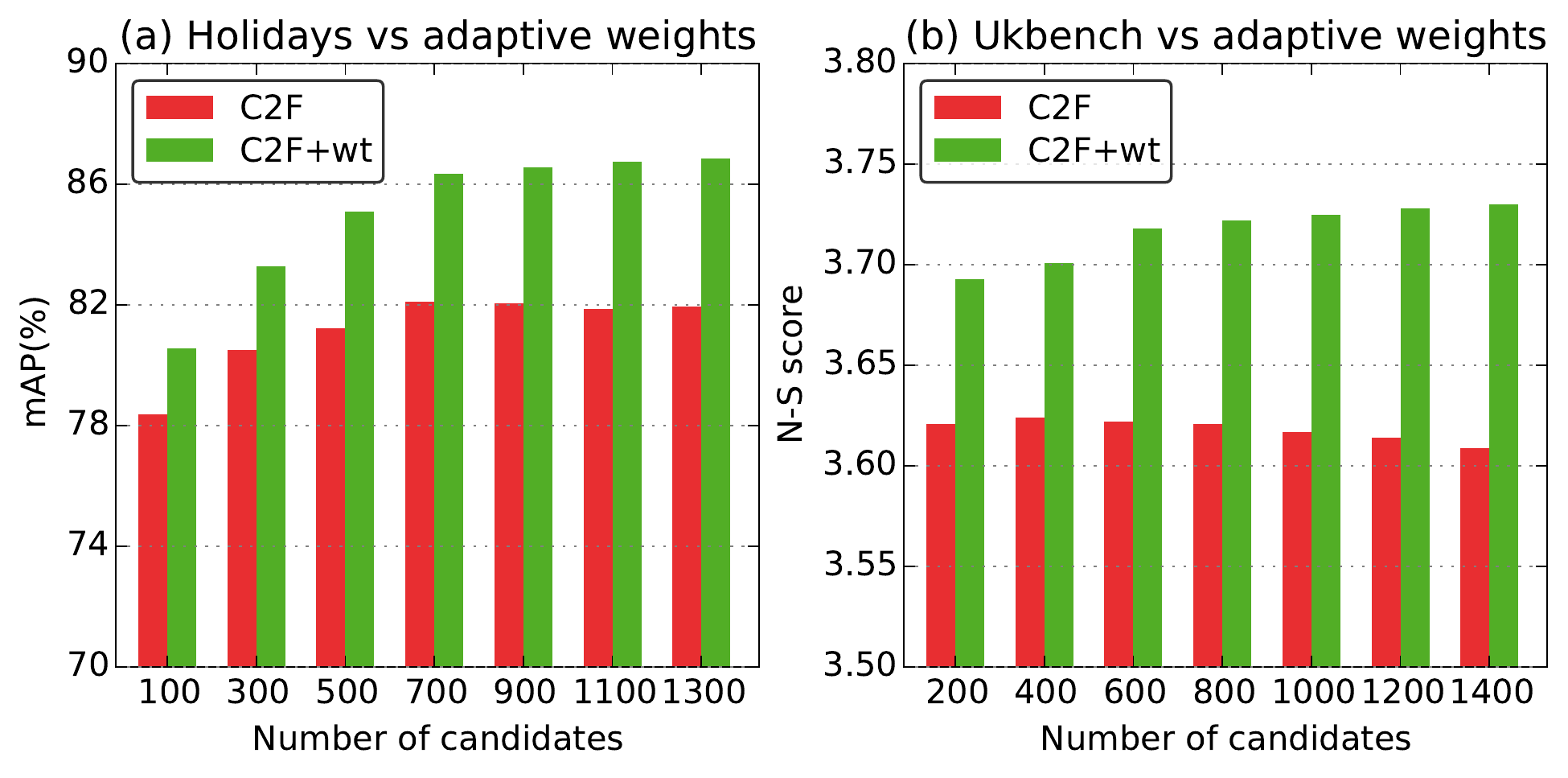}
\end{center}
\caption{The impact of adaptive weights. We compare the performance with the case where no weight is used. The red bar represents the C2F result with\-out weights, while green bar shows results with adaptive weights.}
\label{Fig: weights}
\end{figure}

The Figure \ref{Fig: candidates}(b) shows that candidate-1600 achieves desirable accuracy (N-S = 3.733). With the further increase of the candidates, the retrieval performance shows subtle improvement. It means that when $K$ = 1600, ground truths of most queries are recalled via holistic representation, and more candidate images can not significantly enhance the retrieval precision. What is more, C2F without any weights achieves the best accuracy when candidate number $K$ is set to 800 (N-S = 3.629), and the retrieval performance declines with the further increase of $K$. For example, candidate images with 1400 and 1600 achieve the N-S scores of 3.601 and 3.603, which are 0.028 and 0.025 lower than that of 800. All of the above analysis demonstrates that the set of candidate number plays some dominated role in C2F.

\subsubsection{Evaluation on the adaptive weights} \label{exper: weights}

To demonstrate the effectiveness of the adaptive weights, we compare with the case in which no weights are used. In other words, holistic representation is only used to filter out noisy images and the selected $K$ candidates are considered to be equally important. The results are shown in Figure \ref{Fig: weights}. It is clear that, the usage of adaptive weighs brings benefits for accurate image retrieval.

Compared with the C2F without weights, adaptive weights help to achieve desirable retrieval performance on holidays dataset, as shown in Figure \ref{Fig: weights}(a). When the candidate number $K$ is set to 500, the mAP of C2F with weights is 3.88\% higher than that of without weights. And this gap will enhance with the increase of candidate images. Take the retrieval result with candidate number $K$ = 1400 as an example, the accuracy of C2F with weights is 85.93, which is 4.70\% higher than that of without weights.

From the Figure \ref{Fig: weights}(b), we can obviously observe that, on Ukbench dataset, C2F with adaptive weights significantly outperforms that without weights. When the candidate number is set to 1000, the N-S score of C2F with weights is 3.725, which is 0.108 higher than that without weights. Obviously, with the increase of candidate images, the advantage of the adaptive weights is always exist. In general, the retrieval results show that the weighted strategy significantly influences the final retrieval performance when all the other settings keep the same.

\subsubsection{Comparison with state-of-the-arts} \label{exper: state-of-the-arts}

Table \ref{table: arts} shows that our method yields competitive results when compared with state-of-the-arts. On holidays dataset, C2F with adaptive weights achieves desirable performance among these methods. With the same holistic and local descriptors, \emph{e.g.}, HSV histogram with 1000-dimensional and BOW embedded with hamming signature, C2F achieves the accuracy of 86.78, which is 3.31\% higher than the query-adaptive fusion method \cite{zheng2015query}. Meanwhile, we observe that our result on Ukbench is slightly lower than \cite{zheng2015query} and \cite{zhang2012query} by 0.013 and 0.018, respectively. This is because the BOW result used in \cite{zheng2015query} is 0.098 higher than ours,\emph{ e.g.}, the mAP of BOW is 3.582 in \cite{zheng2015query}, while in C2F is 3.484. Compared with the graph based method in \cite{zhang2012query}, we use a small loss of accuracy for huge storage saving and efficiency enhancing.

\begin{table}
\centering
\caption{The retrieval performance of C2F with different number of distractor images. Especially, we set the candidate number to 1000, and use the adaptive weights to enhance the retrieval performance.}
\label{table: large-scale}
\begin{tabular}{|c|c|c|c|} \hline
$K$ = 1000 & C2F &C2F+wt \\ \hline
0   & 82.25 & 86.78 \\ \hline
1000 & 80.23 & 83.95 \\ \hline
10,000 & 77.36 & 79.30 \\ \hline
100,000 & 71.64 & 72.73 \\ \hline
500,000 & 68.39 & 69.41 \\ \hline
1,000,000 &67.26 & 69.15 \\ \hline
\end{tabular}
\end{table}

\subsubsection{Large-scale image retrieval analysis} \label{exper: large-scale}

To further evaluate the performance of C2F on the large-scale image datasets, we merge Holidays with ImageNet 1$M$. In the experiments, We use different numbers of distractors to test the scalability of the proposed two-layer fusion framework. Specifically, we set the number of candidates $K$ to 1000, and adopt adaptive weights to enhance the retrieval accuracy.

Table \ref{table: large-scale} shows the retrieval results with different numbers of distractor images. Undoubtedly, the performance of the performance of C2F gradually decreases with the augment of distractors in the database. When the number of distractor images is 0, C2F obtain best accuracy. While when all 1M distractors of ImageNet are added into the Holidays database, the mAP of C2F reduce significantly, which is 12.63\% lower than that of without distractors. Moreover, we do not enhance the corresponing candidates when the distractors increase, so C2F can significantly enhance the efficiency and reduce the memory consumption at the cost of some accuracy loss.

\subsection{Efficiency and Storage} \label{exper: efficiency}

It is well known that BOW incorporated with Hamming Embedding can achieve desirable performance. Meanwhile, a major bottleneck in BOW based methods is the length of the inverted list for it grows almost linearly with the database size. A traverse on the inverted index generated by the whole database can be extravagantly expensive. This problem appears more severe if we consider a typical case where the benchmark contains millions of images, for the inverted list can be as long as GBs. It means that for each query feature, a naive full comparision with all images of the dataset takes considerable time. Table \ref{table: memory} and Table \ref{table: efficiency} give the average query memory consumption and the corresponding time on holidays and Ukbench datasets with the increasing number of candidates, respectively.

With the same candidate number, memory expended on Ukbench is far less than that on Holidays due to the different size of images, while incremental tendency reflected on two benchmarks is consistent. Similarly, from Table \ref{table: memory} and Table \ref{table: efficiency}, we observe that the average query time also significantly increases with the growing number of candidates. This is because that more candidates means larger inverted index, namely, more search time to one query. Thus, an effective filter mechanism is very necessary for large-scale image retrieval, and a moderate $K$ candidate set is important to guarantee the tradeoff between the retrieval accuracy and memory consumption.

\subsection{Discussion} \label{exper: discuss}

\begin{figure}[!htbp]
\begin{center}
\includegraphics[width=1.0\linewidth]{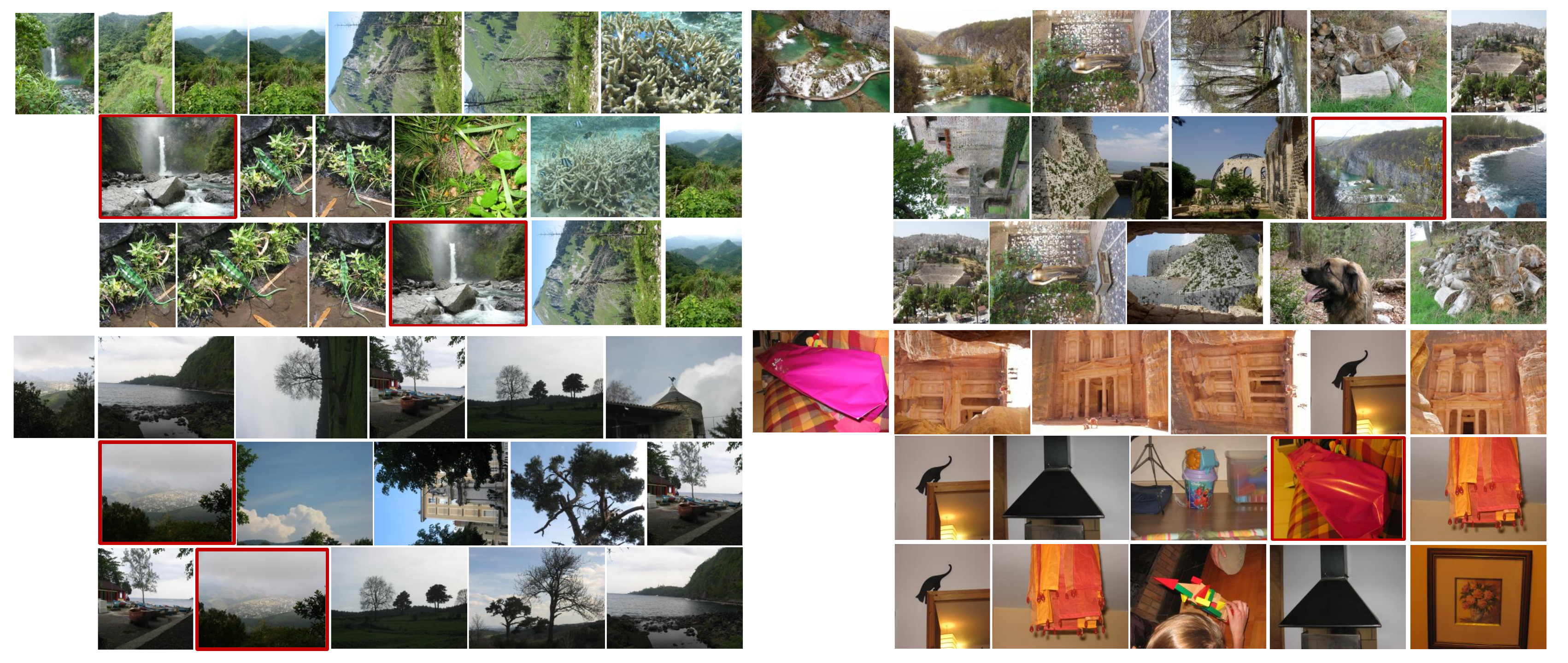}
\end{center}
\caption{Visualization of false retrieval results from Holidays. To each query, top returned images are obtained via HSV histograms (first row), BOW (second row) and the C2F (third row), reapectively. }
\label{Fig: failure}
\end{figure}

Figure \ref{Fig: failure} presents some false retrieval results through C2F on Holidays dataset. As shown in Figure \ref{Fig: failure}, these inaccurate images have the following two characteristics: 1) The image itself contains ambiguous information. Taking the query in the upper left corner as an example, we want the returned top images contain the waterfall, while in fact, the query involves abundant greenery Information. This is why neither global representation nor local feature can acquire satisfactory results. 2) In the experiments, C2F utilizes HSV histograms delineating overall color feature distributions in images to filter out distractors, so the top ranked candidates share similar color composition. This similar error will affect the query accuracy in the next stage, see the queries on the upper right corner and the lower left corner of the Figure \ref{Fig: failure}.

\section{Conclusion}
In this paper, a two-layer fusion method is proposed, which takes advantage of global and local cues and ranks database images from coarse to fine. The main purpose of C2F is to reduce memory consumption and the computational complexity, without compromising the retrieval accuracy. To achieve this goal, C2F adopts holistic representation to filter out noisy images of the benchmark and choose the images with high similarity scores as candidates. Particularly, for each candidate, an adaptive weight is learned via the holistic similarity scores. Then retrieval is conducted on candidate set by taking the adaptive weights into account. Comprehensive experiments are conducted to evaluate the accuracy and scalability of the C2F. With the same holistic and local descriptors, the accuracy of C2F on Holidays is 3.31\% higher than the query-adaptive fusion method \cite{zheng2015query}. In future work, we will further explore more effective holistic representations and design more adaptive weighted functions.


%

%

%
%

\ifCLASSOPTIONcaptionsoff
  \newpage
\fi

\end{document}